\documentclass[preprint,3p,12pt]{elsarticle}

\usepackage{lineno,hyperref}
\modulolinenumbers[5]

\usepackage{hyperref}

\usepackage{epstopdf}

\usepackage{mathrsfs}
\usepackage{amssymb}
\usepackage{amsfonts}
\usepackage{latexsym}
\usepackage{amsmath}
\usepackage{xcolor}
\usepackage{wrapfig}
\usepackage{floatflt}

\usepackage{graphicx}
\usepackage{subcaption}
\def\lb{\label}

\newcommand{\er}[1]{\textrm{(\ref{#1})}}

\newtheorem{theorem}{\bf Theorem}[section]
\newtheorem{lemma}[theorem]{\bf Lemma}

\def\a{\alpha}  \def\cA{{\mathcal A}}       \def\mA{{\mathscr A}}
   \def\cB{{\mathcal B}}

\def\d{\delta}         
  \def\cF{{\mathcal F}}       
    \def\cG{{\mathcal G}}

    \def\cJ{{\mathcal J}}       
        
         \def\mL{{\mathscr L}}
\def\l{\lambda}        
        
\def\m{\mu}

\def\s{\sigma}  \def\cR{{\mathcal R}}

\def\ve{\varepsilon}       \def\vp{\varphi}    

\def\Z{{\mathbb Z}}    \def\R{{\mathbb R}}   \def\C{{\mathbb C}}    
    \def\N{{\mathbb N}}   

\def\lt{\biggl}                  \def\rt{\biggr}
               \def\wt{\widetilde}

\let\ge\geqslant                 \let\le\leqslant

\def\iy{\infty}
\def\sm{\setminus}               
\def\ss{\subset}                 
                
                 \def\ev{\equiv}
        
\def\el2{\ell^{\,2}}             \def\1{1\!\!1}

\def\const{\mathop{\mathrm{const}}\nolimits}

\def\diag{\mathop{\mathrm{diag}}\nolimits}

\def\dim{\mathop{\mathrm{dim}}\nolimits}

\def\Res{\mathop{\mathrm{Res}}\nolimits}

\def\BBox{\hspace{1mm}\vrule height6pt width5.5pt depth0pt \hspace{6pt}}

\newtheorem{corollary}[theorem]{\bf Corollary}

\let\ge\geqslant
\let\le\leqslant

\newcommand{\ca}{\begin{cases}}
\newcommand{\ac}{\end{cases}}
\newcommand{\ma}{\begin{pmatrix}}
\newcommand{\am}{\end{pmatrix}}
\def\eq{\begin{equation}}
\def\qe{\end{equation}}
\renewcommand{\[}{\begin{equation}}
\renewcommand{\]}{\end{equation}}

\def\BBox{\hspace{1mm}\vrule height6pt width5.5pt depth0pt \hspace{6pt}}

\begin{document}

\begin{frontmatter}

\title{On some explicit integrals related to ``fractal mountains''}
\date{\today}

\author
{Anton A. Kutsenko}

\address{Jacobs University (International University Bremen), 28759 Bremen, Germany; email: akucenko@gmail.com}

\begin{abstract}
Loop counting functions $U(x)$ estimate the  number of ``weighted" loops in a digital representation of $x\in[-1,1]$. Roughly speaking, each $x$ is considered as an infinite walk, where the steps of the walk correspond to digits of $x$. The graph of loop counting functions $U$ has a fractal structure that resembles complex mountain landscapes. In some sense, $U$ allows us to look at random walks globally. These functions may be helpful in the analysis of some hard problems related to the distribution of self-avoiding random walks (SAW) in a multi-dimensional case since SAW closely relate to zeros of $U(x)$. We note here that $U(x)$ can be naturally extended to a multidimensional argument $x$. In this article, the focus will be on some analytic aspects. It will be shown that integrals $\int x^AU(x)^Bdx$ with non-negative integers $A$ and $B$ can be expressed in terms of integrals of rational functions with integer coefficients. Moreover, it will be shown that $\int x^A U(x)dx$ admits closed-form expressions. Fourier series for $U$ is also computed. Finally, we discuss some connections with special functions and generalized continued fractions, and other perspectives. 
\end{abstract}



\end{frontmatter}


{\section{Introduction}\lb{sec1}}

Loop counting functions can be considered as a variation of Takagi curves, because both use dyadic expansions of numbers in their construction. The definition of such curves is given in \cite{T} and some recent results are discussed in \cite{MS}. However, the structure of Takagi curves and their purpose are significantly different from that of loop counting functions. Takagi curves are continuous and they illustrate simple examples of curves that have no derivatives everywhere. Loop counting functions $U(x)$ relates to loops in random walks that represented by real values $x$. They are discontinuous. One of the purposes of theses functions is an attempt to study self-avoiding random walks (SAW), since, roughly speaking, $U(x)=0$ if and only if $x$ corresponds to SAW. There are some open problems related to SAW, see, e.g., \cite{S}. Another purpose of loop counting curves is the expository objective. The structure of these curves allows to see globally some characteristics of a distribution of loops in random walks that represented by real numbers. The representation of concrete real numbers ($\pi$, $\ln2$, ...) by random walks is well illustrated in a nice paper \cite{ABBB}. The corresponding approach aims to understand the normality of specific numbers that is a hard open problem in number theory. In this research, I mostly focus on different problems related to SAW indirectly. Namely, we will try to compute various integrals of functions depended on $U$. I believe that in the future these results can help to analyze the distribution of zeros of $U$ in a multidimensional case. But at the moment our goals more prosaic: while $U$ has a complex fractal structure, it will be shown that various integrals of $U$ can be reduced to the integrals of rational functions with integer coefficients. Further, I will plan to find the representation of the algebras of functions acting on the loops of random walks and on other fractal structures that uses dyadic representations in the same way as it was made in \cite{K1} and \cite{K2}.

Any $x\in[-1,1]$ except a countable set of some dyadic rationals can be uniquely expanded as
\[\lb{001}
 x=\frac{x_0}2+\frac{x_1}{2^2}+\frac{x_2}{2^3}+...,\ \ x_n\in\{-1,+1\}.
\]
For $0\le n\le m$, define the functions that determine loops
\[\lb{002}
 L_{nm}(x)=\ca 1,& \sum_{j=n}^{m}x_j=0,\\
               0,& otherwise. \ac
\]
Let us fix $\l$ such that $|\l|<1$. Define the function that count weighted loops
\[\lb{003}
 U(x)=\sum_{0\le n\le m<+\iy}\l^{m+1}L_{nm}(x).
\]
Function $U$ is even $U(x,\l)=U(-x,\l)$. The function has a fractal structure, see Fig. \ref{fig1}. There is some self-similarity, but if we zoom some elements of the curve, they look differently from segment to segment, see Fig. \ref{fig2}. 
\begin{figure}
  \centering
    \includegraphics[width=0.99\textwidth]{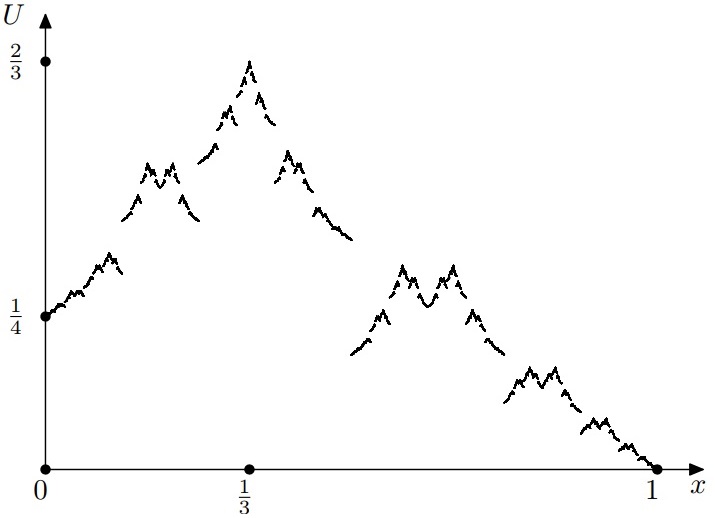}
  \caption{The plot of $U$ for $\l=1/2$. \lb{fig1}}  
\end{figure}
\begin{figure}
    \centering
    \begin{subfigure}[b]{0.49\textwidth}
        \includegraphics[width=\textwidth]{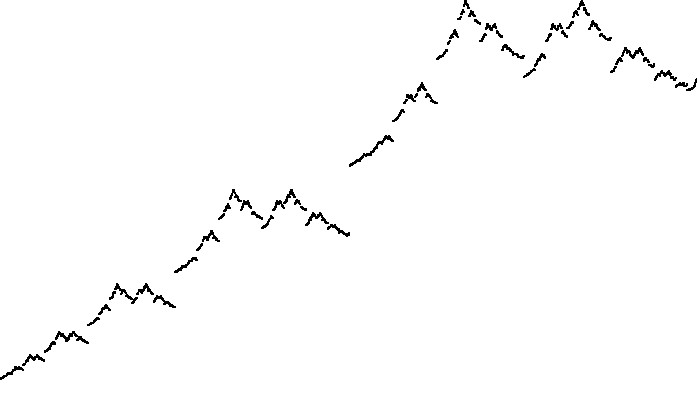}
    \end{subfigure}
    \begin{subfigure}[b]{0.49\textwidth}
        \includegraphics[width=\textwidth]{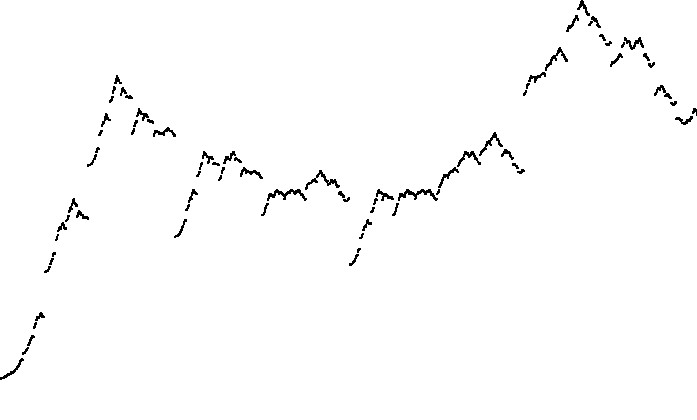}
    \end{subfigure}
    \caption{For $\l=1/2$, two randomly chosen different segments of the curve $U$ zoomed in $\approx2^{70}$ times. 
}\label{fig2}
\end{figure}
Following the results of \cite{K3}, we can compute the integral of $U$ explicitly
\begin{multline}\lb{L1norm}
	\int_0^1U(x)dx=\sum_{k=1}^{+\iy}\sum_{m,n\ge0,\ m-n=2k-1}4^{-k}\binom{2k}{k}\l^{m+1}=\sum_{k=1}^{+\iy}4^{-k}\binom{2k}{k}\sum_{m=2k-1}^{+\iy}\l^{m+1}=\\
	\sum_{k=1}^{+\iy}4^{-k}\binom{2k}{k}\frac{\l^{2k}}{1-\l}=\frac{1}{(1-\l)\sqrt{1-\l^2}}-\frac{1}{1-\l}=\frac{1-\sqrt{1-\l^2}}{(1-\l)\sqrt{1-\l^2}}.
\end{multline}
Note that the results of \cite{K3} dealt with non-negative weights, i.e. $\l\in[0,1)$, but the same ideas can be applied for any $|\l|<1$. While \cite{K3} devoted to general non-negative weights, in the current work I focus on exponential weights $\l^n$. This allows us to compute explicitly more complex integrals $\int x^AU(x)^Bdx$ than \er{L1norm} by using different methods, e.g., the analysis of functions of several complex variables. Let us not that exponential weight $\l^n$ with $|\l|<1$ decay to $0$ so fast that checking the convergence, the possibility of changing the order of summation and integration, and other similar conditions become trivial. We will often omit the corresponding reasoning.
\begin{theorem}\lb{T1}
Let $A,B,N\ge0$ and $-2^N<c<2^N$ be integer numbers.
Then there exist polynomials $P$, $Q$ of $\l,z_1,...,z_B$ with integer coefficients such that
\[\lb{004}
\int_{\frac{c-1}{2^N}}^{\frac{c+1}{2^N}}x^AU(x)^Bdx=\frac1{(2\pi i)^B}\oint_{|z_1|=1}...\oint_{|z_B|=1}\frac{P(\l,z_1,...,z_B)}{Q(\l,z_1,...,z_B)}dz_B...dz_1.
\]
The polynomials $P$, $Q$ can be constructed explicitly. For example, up to a polynomial of $\l$ and simple power factors of $z_i$ the polynomial $Q$ is a product of multivariate quadratic polynomials of the form $\l^m\prod_{i\in\a}z_i^2-2^{n+1}\prod_{i\in\a}z_i+\l^m$, where set $\a\ss\{1,...,B\}$, and $0\le n\le A$, $0\le m\le B$.
\end{theorem}
The algorithm of construction of $P$, $Q$ is given in the Section devoted to the proof of the theorem. It should be noted that zeros of $Q$ can be computed explicitly. This allows us to apply the Cauchy residue theorem for obtaining closed-form expressions. We consider the simplest case $B=1$. 
\begin{corollary}\lb{C1} Under the conditions of Theorem \ref{T1}, integrals $\int_{\frac{c-1}{2^N}}^{\frac{c+1}{2^N}}x^AU(x)dx$ admit closed form expressions through rational functions of $\l$ and square roots of them.
\end{corollary}
This corollary extends \er{L1norm}. Let us provide one example of explicit computation $\int_{-1}^1U(x)x^Adx$, $A\ge0$ - in fact, the non-trivial cases correspond to even $A$. For $n\ge0$, define the quadratic polynomials 
$$
 Q_n(z)=2^{n+1}z-\l z^2-\l
$$
and the following rational functions
$$
 f_n(z)=\frac{2^{n+1}(1+(-1)^n)}{(n+1)Q_n(z)}+\sum_{j=1}^n\sum_{n=n_0>n_1>...>n_j\ge0}\frac{2^{n_j+1}(1+(-1)^{n_j})}{(n_j+1)Q_{n_j}(z)}\prod_{i=1}^j\frac{\l\binom{n_{i-1}}{n_i}(z^2+(-1)^{n_{i-1}-n_i})}{Q_{n_{i-1}}(z)},
$$
where $\sum_{j=1}^0...=0$ is assumed.
Now let us define define the quantities
$$
 \hat f_n=\sum_{j=0}^n\Res(f_n,z_j),\ \ \ z_j=\frac{\l}{2^j+\sqrt{4^j-\l}}\ \ (|z_j|<1)
$$
that are, in fact, integrals $(2\pi i)^{-1}\oint f_n(z)dz$ over the unit circle, since $z_j$ are zeros of $Q_j(z)$, i.e. $z_j$ are all the possible poles of $f_n$. It is seen that $\hat f_n$ can be expressed through rational functions of $\l$ and square roots of them. Finally, using \er{H023} along with \er{H022}, \er{H020}, \er{H016} and \er{H014}, it is possible to express the integral $\int_{-1}^1U(x)x^Adx$ through $\hat f_n$ by
\begin{multline}
 \int_{-1}^1x^AU(x)dx=\frac{2^A((A+1)\hat f_A-1-(-1)^A)}{(A+1)(2^A-\l)}+\\
 \sum_{j=1}^A\sum_{A=n_0>n_1>...>n_j\ge0}\frac{2^{n_j}((n_j+1)\hat f_{n_j}-1-(-1)^{n_j})}{(n_j+1)(2^{n_j}-\l)}\prod_{i=1}^j\frac{\l\binom{n_{i-1}}{n_i}(1+(-1)^{n_{i-1}-n_i})}{2^{n_{i-1}+1}-2\l}.\notag
\end{multline}
As an example, I provide the following integral, see \er{H028},
$$
 \int_0^1x^2U(x)dx=\frac{1}{3(1-\l)}\lt(\frac1{\sqrt{1-\l^2}}-1\rt)+\frac{8\sqrt{1-\l^2}-12\sqrt{4-\l^2}+4\sqrt{16-\l^2}}{3(4-\l)}.
$$
For $\l=1/2$, both numerical integration and analytic result give $\int_0^1x^2U(x)dx\approx0.1234/2$, but the numerical integration requires much more computations than the analytic formula. 

Integrals \er{004} can also be computed for $B>1$ by using Cauchy residue theorem. In this article, I do not discuss the closed-form expressions for the case $B>1$. I provide only the formula for the second moment, see \er{245},
$$
 \int_0^1U(x)^2dx=\frac1{(1-\l^2)(1-\l)^2}\lt(\frac{(1-\l)^2}{\sqrt{1-\l^4}}+\frac{4\l}{1+(1+\l)\sqrt{1+\l^2}}+1-\l^2-2\sqrt{1-\l^2}\rt).
$$
For $\l=1/2$, both numerical integration and analytic result give $\int_0^1U(x)^2dx\approx0.2479/2$, but again the numerical integration requires much more computations than the analytic formula.

Let us discuss briefly perspectives related to the function $U$ in view of functional analysis . Define the following operator (functional)
\[\lb{005}
 \cG f=\int_{-1}^{1}U(x)f(x)dx.
\] 
Since $U$ is a bounded function, $\cG$ can be considered as a bounded operator, say $\cG:C([-1,1]\to\C)\to\C$ or $\cG:L^2([-1,1]\to\C)\to\C$. Let us express $\cG$ through some more simple operators. Define one functional and two operators
\[\lb{006}
 \cJ f=\int_{-1}^1f(x)dx,\ \ \ (\cR_{-}f)(x)=f\lt(\frac{x-1}2\rt),\ \ \ (\cR_{+}f)(x)=f\lt(\frac{x+1}2\rt).
\]
Again, all these operators and the functional are bounded on both spaces: the space of continuous functions and square-integrable functions. 
\begin{theorem}\lb{T2}
The following identity holds
\[\lb{007}
 \cG=\l\cJ\oint_{|z|=1}\lt(1-\l\frac{z\cR_++z^{-1}\cR_-}2\rt)^{-1}(z\cR_++z^{-1}\cR_-)\frac{dz}{4\pi i z}\lt(1-\l\frac{\cR_++\cR_-}2\rt)^{-1}.
\]
\end{theorem}
For $|\l|<1$ the inverse operators can be expanded into the standard geometric series, since the norm $\|z\cR_++z^{-1}\cR_-\|\le 2$ in both spaces $C$ and $L^2$ as it is shown in the proof of Theorem \ref{T2}. Moreover, the operators $\cR_{\pm}$ have lower triangular matrix representations in the polynomial basis $\{x^n\}_{n\ge0}$. This means that the inverse operators can be computed explicitly. Thus, in the polynomial basis the integrand in \er{007} can also be computed explicitly. After that, Cauchy residue theorem  allows as to compute $\cG$. In details, this step is explained above and in the example Section \ref{sec3}. As another application of Theorem \ref{T2}, I formulate the following result.

\begin{figure}
    \centering
    \begin{subfigure}[b]{0.49\textwidth}
        \includegraphics[width=\textwidth]{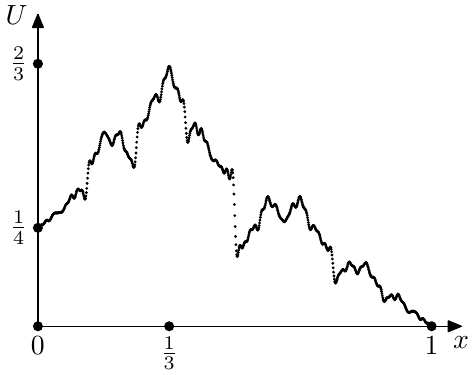}
        \caption{$150$ harmonics}
    \end{subfigure}  
    \begin{subfigure}[b]{0.49\textwidth}
        \includegraphics[width=\textwidth]{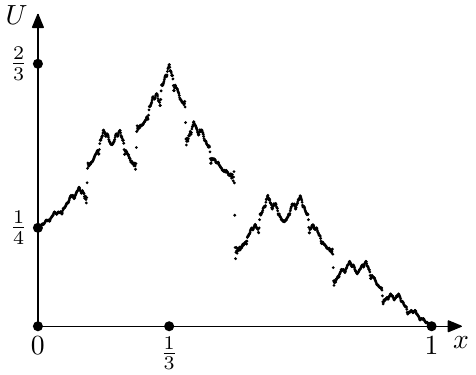}
        \caption{$500$ harmonics}
    \end{subfigure} 
    \caption{For $\l=1/2$, the approximation of $U$ in the trigonometric Fourier basis $\{\cos n\pi x\}_{n\ge0}$ is plotted.}\label{fig3}
\end{figure}

\begin{corollary}\lb{C2} For $\s\in\C\sm\pi\Z$, the following identity holds
\[\lb{008}
 \int_{0}^1U(x)\cos\s xdx=\frac{\sin\s}{\s}\sum_{n=0}^{\iy}\frac{\l^n}{\sin\frac{\s}{2^n}}\sum_{m=1}^{\iy}\l^mC_m(\frac{\s}{2^n})\sin\frac{\s}{2^{n+m}},\ \ where
\]
\begin{multline}\lb{009}
	C_m(\s)=2^m\int_{-\pi}^{\pi}\prod_{j=1}^m\cos(\vp+\frac{\s}{2^j})\frac{d\vp}{2\pi}=\sum_{{\bf y}\in\{-1,1\}^m_0}\cos{\sum_{j=1}^m\frac{\s y_j}{2^j}},\ \ \ with\\ 
	\{-1,1\}^m_0=\{{\bf y}=(y_j)_{j=1}^m\in\{-1,1\}^m:\ \sum_{j=1}^my_j=0\}.
\end{multline}
Note that $C_m\ev0$ for odd $m$. Now, consider $\s\in\pi\Z$. The case $\s=0$ is already given in \er{L1norm}. If $\s=\pi 2^pq$, $p\in\N\cup\{0\}$, $q\in2\Z+1$ then the double sum in \er{009} can be reduced to one
\begin{multline}\lb{009a}
 \int_{0}^1U(x)\cos\pi 2^pq xdx= \frac{\l^p(-1)^{\d_{p0}}}{\pi q}	\sum_{n=0}^{p}(-1)^{\d_{n0}}2^{-n}\sum_{m=1}^{\iy}\l^{m}C_{m+n}(\pi2^nq)\sin\frac{\pi q}{2^m}=\\
 \frac{(-1)^{1-\d_{p0}}\l^p}{\pi q}\sum_{m=1}^{\iy}(2\l)^m\sin\frac{\pi q}{2^m}\int_{-\pi}^{\pi}\frac{(1-\cos^{p+1}\vp)\prod_{j=1}^{m}\cos(\vp+\frac{\pi q}{2^j})}{1-\cos\vp}\frac{d\vp}{2\pi}.
\end{multline}
\end{corollary}
While \er{008} expresses $\int_{0}^1U(x)\cos\s xdx$ in terms of standard functions, this is not a closed-form expression in the usual sense, since \er{008} includes infinite series. Both series converges sufficiently fast. The first formula in \er{009} allows us to develop some convenient computational procedure. Let us start from the case $\s\ne\pi k$, $k\in\Z$. Define two functions
\[\lb{010}
 C(\vp,\s)=\sum_{m=1}^{\iy}\prod_{j=1}^m\frac{\l\cos(\vp+\frac{\s}{2^j})}{\cos\frac{\s}{2^j}},\ \ \ D(\vp,\s)=\sum_{n=0}^{\iy}\l^nC(\vp,\frac{\s}{2^n}).
\]
Again, it is assumed that $|\l|<1$. Using \er{008}, the first formula in \er{009}, and 
identity $\sin\frac{\s}{2^n}=2^m\sin\frac{\s}{2^{n+m}}\prod_{j=1}^m\cos\frac{\s}{2^{n+j}}$, we deduce that
\[\lb{011}
  \int_{0}^1U(x)\cos\s xdx=\frac{\sin\s}{ \s}\int_{-\pi}^{\pi}D(\vp,\s)\frac{d\vp}{2\pi}.
\] 
Functions \er{010} satisfy recurrence relations
\[\lb{012}
 C(\vp,\s)=\frac{\l\cos(\vp+\frac{\s}{2})}{\cos\frac{\s}{2}}(1+C(\vp,\frac{\s}2)), \ \ \ D(\vp,\s)=C(\vp,\s)+\l D(\vp,\frac{\s}2).
\]
The numerical procedure consists of choosing some large number $N$ such that $\s/2^N\approx0$ and taking
$$
 \wt C_N=C(\vp,0)=\frac{\l\cos\vp}{1-\l\cos\vp},\ \ \ \wt D_N:=D(\vp,0)=\frac{C(\vp,0)}{1-\l}.
$$
After that the recurrence formulas \er{012} can be applied in the following form
$$
 \wt C_{n-1}=\frac{\l\cos(\vp+\frac{\s}{2^n})}{\cos\frac{\s}{2^n}}(1+\wt C_n),\ \ \ \wt D_{n-1}=\wt C_{n-1}+\l\wt D_n.
$$
Hence, we use a backward scheme. While a forward scheme is more traditional, it does not give a real advantage in this case. At the end, $\wt D_1$ gives a good approximation of $D(\vp,\s)$. It is obvious that for $\vp\in\R$ and any fixed $|\l|<1$, $\s\in\C\sm\pi\Z$, the convergence of the scheme is uniformly exponentially fast with the rate $\le\const(\s)|\l|^N$. Repeating these steps for $\vp=\pi r/M$, $r=-M+1,...,M$ for some large $M$ and then taking the Riemann sum approximating second integral in \er{011}, we obtain a good approximation of $\int_{0}^1U(x)\cos\s xdx$. I have checked this scheme and found that often it looks more efficient than the direct numerical computation of $\int_{0}^1U(x)\cos\s xdx$, since the scheme dealt with analytic functions. Finally, note that if $\s=2^pq\pi$, under the same conditions as in \er{009a}, then
\begin{multline}\lb{013}
 \int_{0}^1U(x)\cos\pi 2^pq xdx= \\
 \frac{(-1)^{1-\d_{p0}}2\l^{p+1}\sin\frac{\pi q}{2}}{\pi q}
 \int_{-\pi}^{\pi}\frac{(1-\cos^{p+1}\vp)\cos(\vp+\frac{\pi q}{2})C(\vp,\frac{\pi q}2)}{1-\cos\vp}\frac{d\vp}{2\pi},
\end{multline}
where the fact that $\int_{-\pi}^{\pi}A(\vp)\cos(\vp+\frac{\pi q}{2})d\vp=0$ for any even function $A(\vp)=A(-\vp)$ is used. Thus, the analog of numerical scheme for $\s=2^pq\pi$ is even simpler than in the general case considered above. In Fig. \ref{fig3}, I plot approximations of $U$ based on the trigonometric Fourier series with the coefficients computed by \er{013} and \er{L1norm} for the free term. Further simplifications may based on the properties of $C$ and on the fact that $C_{2m+1}\ev0$, see above. 
For example, applying the Euler continued fraction formula and an equivalence transformation to the first identity in \er{010}, we obtain
\begin{multline}\lb{014}
C(\vp,\s)=\\
\cfrac{\l\cos(\vp+\frac{\s}2)}{{\cos\frac{\s}2}-\cfrac{\l{\cos\frac{\s}2}\cos(\vp+\frac{\s}4)}{ {\cos\frac{\s}4}+\l\cos(\vp+\frac{\s}4)-\cfrac{\l{\cos\frac{\s}4}\cos(\vp+\frac{\s}8)}{{\cos\frac{\s}8}+\l\cos(\vp+\frac{\s}8)-
			\cfrac{\l{\cos\frac{\s}8}\cos(\vp+\frac{\s}{16})}{{\cos\frac{\s}{16}}+\l\cos(\vp+\frac{\s}{16})-\cdots}
		}}}.
\end{multline}
At the same time, it is not clear to me which expression for $C$ maybe helpful in the further analysis of Fourier coefficients of $U$. It may happens that Fourier coefficients in the Legendre polynomial basis are more perspective for the analysis than those in the trigonometric basis, since Fourier coefficients in the Legendre polynomial system admit finite closed-form expressions. Any choice depend on our goals. We plan to study distribution of small $U$ values in the multidimensional case. For this, we have firstly to develop various techniques on the one-dimensional case. However, despite all this, the main motivation remains the search for interesting formulas related to $U$.

Further work is organized as follows: in Section \ref{sec2} I prove Theorems \ref{T1} and \ref{T2}; in Section \ref{sec3} I prove Corollary \ref{C1}, discuss integrals $\int \frac{U(x)}{w-x}$, and construct explicitly rational functions that computes $\int U(x)^2$; in Section \ref{sec4} I discuss various perspectives, including important multivariate loop counting functions $U(x_1,...,x_d)$. Other explicit integrals of some functions similar to $U(x)$ but with more complex loop's weights than $\l^n$ are also discussed in Section \ref{sec4}. Finally, some linear operators acting in functional spaces related to loop-counting functions are  discussed briefly.

{\section{Proof of Theorem \ref{T1}, \ref{T2}, and Corollary \ref{C2}}\lb{sec2}}

For $x\in[-1,1]$, let us define the functions
\[\lb{101}
 F(x,z)=1+\l z^{x_0}+\l^2z^{x_0+x_1}+\l^3z^{x_0+x_1+x_2}+...=1+\sum_{n=0}^{+\iy}\l^{n+1}z^{\sum_{j=0}^nx_j},
\]
\[\lb{102}
 G(x,z)=\sum_{0\le n\le m<+\iy}\l^{m+1}z^{\sum_{j=n}^mx_j},
\]
where $x_n\in\{-1,+1\}$ are given by \er{001}. Since $|\l|<1$, it is seen that for any $x\in[-1,1]$ the functions $F$ and $G$ are analytic in some open ring containing the circle $|z|=1$. Indeed, each term of the series \er{101} can be uniformly approximated by the terms of convergent series
\[\lb{c001}
 |F(x,z)|\le 1+\l\max\{|z|,|z|^{-1}\}+\l^2\max\{|z|^2,|z|^{-2}\}+...\le\frac1{1-\l|z|}+\frac1{1-\l|z|^{-1}},
\]
since all $x_n\in\{-1,+1\}$. Thus $F(x,z)$ is analytic in $\{z:\ \l<|z|<\l^{-1}\}$ for any fixed $x\in[-1,1]$. Similar arguments work for $G$ too. Moreover, both functions are measurable as functions of two arguments, since they can be approximated by step functions depended on $x$ multiplied by rational functions depended on $z$, in the same way as it is shown in \cite{K3}. The following lemma is basic in our research.
\begin{lemma}\lb{L1}Let us fix some
$$
 \wt x=\frac{\wt x_0}2+...+\frac{\wt x_N}{2^{N+1}},\ \ \wt x_n\in\{-1,+1\}
$$ 
for some $N\ge0$. Then for any $y\in[-2^{-N-1},2^{-N-1}]$ the following identities hold
\[\lb{103}
 F(\wt x+y,z)=1+\l z^{\wt x_0}+...+\l^{N+1}z^{\wt x_0+...+\wt x_N}F(2^{N+1}y,z),
\]
\[\lb{104}
 G(\wt x+y,z)=\sum_{0\le n\le m< N}\l^{m+1}z^{\sum_{j=n}^m\wt x_j}+\l^{N+1}\sum_{n=0}^Nz^{\sum_{j=n}^N\wt x_j}F(2^{N+1}y,z)+\l^{N+1}G(2^{N+1}y,z).
\]
\end{lemma}

{\it Proof.} If $y\in[-2^{-N-1},2^{-N-1}]$ then 
\[\lb{p001}
 y=\frac{y_{N+1}}{2^{N+2}}+\frac{y_{N+2}}{2^{N+3}}+...\ for\ some\ y_n\in\{-1,+1\}.
\]
Thus, using \er{101}, we have
\begin{multline}\lb{p002}
 F(\wt x+y,z)=1+\l z^{\wt x_0}+...+\l^{N+1}z^{\wt x_0+...+\wt x_N}+\l^{N+2}z^{\wt x_0+...+\wt x_N+y_{N+1}}+\\
 \l^{N+3}z^{\wt x_0+...+\wt x_N+y_{N+1}+y_{N+2}}+...=1+\l z^{\wt x_0}+...+\l^{N+1}z^{\wt x_0+...+\wt x_N}+\\
\l^{N+1}z^{\wt x_0+...+\wt x_N}(1+\l z^{y_{N+1}}+\l^2z^{y_{N+1}+y_{N+2}}+...)=\\
1+\l z^{\wt x_0}+...+\l^{N+1}z^{\wt x_0+...+\wt x_N}F(2^{N+1}y,z).
\end{multline}
Similarly, using \er{102}, we obtain
\begin{multline}\lb{p003}
G(\wt x+y,z)=\sum_{0\le n\le m< N}\l^{m+1}z^{\sum_{j=n}^m\wt x_j}+\sum_{0\le n\le N\le m}\l^{m+1}z^{\sum_{j=n}^{N}\wt x_j+\sum_{j=N+1}^my_j}+\sum_{N< n\le m}\l^{m+1}z^{\sum_{j=n}^m y_j}=\\
\sum_{0\le n\le m< N}\l^{m+1}z^{\sum_{j=n}^m\wt x_j}+\l^{N+1}\sum_{n=0}^Nz^{\sum_{j=n}^N\wt x_j}(1+\sum_{k=1}^{+\iy}\l^kz^{\sum_{j=N+1}^{N+k}y_j})+\l^{N+1}G(2^{N+1}y,z)=\\
\sum_{0\le n\le m< N}\l^{m+1}z^{\sum_{j=n}^m\wt x_j}+\l^{N+1}\sum_{n=0}^Nz^{\sum_{j=n}^N\wt x_j}F(2^{N+1}y,z)+\l^{N+1}G(2^{N+1}y,z),
\end{multline}
where, in the first string, we assume that $\sum_{j=N+1}^m...=0$ if $m\le N$.
\BBox

For $N=0$, formulas \er{103}, \er{104} give
\[\lb{105}
 F(\pm\frac12+y,z)=1+\l z^{\pm1}F(2y,z),\ \ \ y\in[-\frac12,\frac12],
\]
\[\lb{106}
 G(\pm\frac12+y,z)=\l z^{\pm1}F(2y,z)+\l G(2y,z),\ \ \ y\in[-\frac12,\frac12].
\]
Using \er{102}, we see that
\[\lb{107}
x^A\prod_{k=1}^BG(x,z_k)=x^AU(y)^B+\sum_{\a_1,...,\a_k} U_{\a_1,...,\a_k}(x)\prod_{k=1}^Bz_k^{\a_k}
\]
for integer $\a_k$ and some functions $U_{\a_1,...,\a_k}$. The crucial point is that the free term in Taylor-Laurent expansion \er{107} is exactly $x^AU(x)^B$. Thus, we have
\[\lb{108}
 \int_{-1}^{1}x^AU(x)^Bdx=\oint_{|z_1|=1}\frac{dz_1}{2\pi iz_1}...\oint_{|z_B|=1}\frac{dz_B}{2\pi iz_B}\int_{-1}^{1}x^A\prod_{k=1}^BG(x,z_k)dx.
\]

Consider the finite-dimensional linear space $\mL\ev\mL_{A,B}$ over the field of rational functions $\mathbb{Q}_B:=\mathbb{Q}(\l,z_1,...,z_B)$ and generated by the following basis
\[\lb{109}
 {\rm Bas}(\mL)=\{x^m\prod_{i=1}^By_i,\ \ y_i=1\ or\ y_i=G_i\ or\ y_i=F_i,\ \ 0\le m\le A\}. 
\]
Formally, this is a linear space of polynomials of $B+1$ variables, where we set the restrictions on the powers of variables. Let us introduce the vector valued norm of the basis elements
\[\lb{110}
 \mathfrak{n}(x^m\prod_{i=1}^By_i)=(m,n_G,n_F),
\]
where $n_G$ and $n_F$ are the quantities of $G$ and $F$ entries in $\prod_{i=1}^By_i$ respectively. We assume that the basis elements are ordered in the lexicographic order
\[\lb{111}
 (a,b,c)<(a',b',c')\ \Leftrightarrow\ (a<a')\lor((a=a')\land(b<b'))\lor((a=a')\land(b=b')\land(c<c')).
\]
Let us introduce the linear operator $\cA:\mL\to\mL$ by
\begin{multline}\lb{112}
 \cA p(x,F_1,...,F_B,G_1,...,G_B)=\\
 \frac12 p(\frac{x+1}2,1+\l z_1F_1,...,1+\l z_BF_B,\l G_1+\l z_1F_1,...,\l G_B+\l z_BF_B)+\\
 \frac12 p(\frac{x-1}2,1+\l z_1^{-1}F_1,...,1+\l z_B^{-1}F_B,\l G_1+\l z_1^{-1}F_1,...,\l G_B+\l z_B^{-1}F_B).
\end{multline}
In fact, $\cA$ is a change-of-variable operator. It is seen that 
\[\lb{112a}
 \cA x^m\prod_{i=1}^By_i=\frac1{2^m}\l^{n_G+n_F}\frac{{\bf z}_F+{\bf z}_F^{-1}}2x^m\prod_{i=1}^By_i+\sum a_jp_j,\ \ \a_j\in\mathbb{Q}_B,\ \ \mathfrak{n}(p_j)<\mathfrak{n}(x^m\prod_{i=1}^By_i),
\]
where $n_G$ and $n_F$ are the quantities of $G$ and $F$ entries in $\prod_{i=1}^By_i$, and
\[\lb{112b}
 {\bf z}_F^s=\prod_{i=1}^Bw_i^s,\ \ w_i=\ca z_i,& y_i=F_i,\\ 1,& otherwise. \ac
\]
Thus, in the basis \er{109}, the operator $\cA$ has a lower triangular form. All the eigenvalues of $\cA$ are $\frac1{2^m}\l^{n_G+n_F}\frac{{\bf z}_F+{\bf z}_F^{-1}}2$. They are all non-zero and hence $\cA$ has inverse $\cA^{-1}$. Now, let us introduce the functional $\cJ:\mA\to\mathbb{Q}_B$ such that
\[\lb{112c}
 \cJ 1=2,\ \ \ \cJ\cA=\cJ.
\]
In fact, if we associate $F_i$ with $F(x,z_i)$ and $G_i$ with $G(x,z_i)$ then we have that $\cJ=\int_{-1}^1$, since \er{112} along with $\cJ=\cJ\cA$ follows from \er{105} and \er{106}. This also justifies the notation $\cJ$ already used in \er{006}. Above, we have seen that $\cA$ has the eigenvalue $1$ with the multiplicity $1$. The corresponding (right) eigenvectors are constants. The condition $\cJ\cA=\cJ$ means that the functional $\cJ$ is the left eigenvector defined uniquely by $\cJ 1=2$. Moreover, since $\cA$ has a lower triangular form in the basis \er{109}, it simplifies the explicit computation of $\cJ$. All components of $\cJ$, namely its values on the basis functions \er{109} are belongs to $\mathbb{Q}_B$. In particular,
\[\lb{112d}
 \mathbb{Q}_B\ni\cJ x^AG_1...G_B=\int_{-1}^1x^AG(x,z_1)...G(x,z_B)dx,
\]
which with \er{108} finishes the proof of \er{004} for $c=0$ and $N=0$.  It remains for us to prove that the essential part of the denominator $Q$ in \er{004} consists of the product of multivariate quadratic polynomials. Indeed, using \er{112}, \er{112a}, we can write \er{112c} in the basis \er{109} as
\[\lb{112e}
 \ma J_1 & J_2 & J_3 & * & 2 \am\ma
   \m_{1} & 0 & 0 & * & 0 \\
    \a_{21} & \m_{2} & 0 & * & 0\\
    \a_{31} & \a_{32} & \m_3 & * & 0 \\
    * & * & * & * & * \\
    \a_{S1} & \a_{S2} & \a_{S3} & * & 1
 \am=\ma J_1 & J_2 & J_3 & * & 2 \am,
\]
where $S=\dim\mL_{A,B}$, the coefficients $\a_{ij}$ have the form $R_{ij}(\l,z_1,...,z_B,z_1^{-1},...,z_B^{-1})$ with some polynomial $R_{ij}$ having rational coefficients. The diagonal elements $\m_i$ are eigenvalues of $\cA$. They have the form $\m=\frac1{2^m}\l^{n_G+n_F}\frac{{\bf z}_F+{\bf z}_F^{-1}}2$ computed above. The norm $\mathfrak{n}(x^AG_1...G_B)$ is maximal among basis elements \er{109}. Hence, RHS of \er{112d} is exactly $J_1$. 
In turn, as it is easy to see, the solution of \er{112e} has the form
\[\lb{112f}
 J_1=\frac{R_1(\{\a_{ij},\m_i\})}{(\m_1-1)...(\m_{S-1}-1)}=\frac{R_2(\l,z_1,...,z_B,\l^{-1},z_1^{-1},...,z_B^{-1})}{(\m_1-1)...(\m_{S-1}-1)},
\]
where $R_1$, $R_2$ are polynomials with rational coefficients, that gives the expansion of the denominator $Q$ mentioned in the formulation of Theorem \ref{T1}. We proved Theorem \ref{T1} for the case $c=0$ and $N=0$. The general case $c,N\ne0$ follows from \er{105} and \er{106}.

Now, let us prove Theorem \ref{T2}. Introduce the following functionals
\[\lb{113}
 \cF(z) f=\int_{-1}^1 F(x,z)f(x)dx,\ \ \ \cG(z) f=\int_{-1}^1 G(x,z)f(x)dx
\]
Using recurrent identities \er{105}, \er{106} and definitions \er{006} we obtain
\begin{multline}\lb{114}
 \int_{-\frac12}^{\frac12}G(\pm\frac12+y,z)f(\pm\frac12+y,z)dy=\l z^{\pm1}\int_{-\frac12}^{\frac12}F(2y,z)f(\pm\frac12+y,z)dy+\\
 \l\int_{-\frac12}^{\frac12}G(2y,z)f(\pm\frac12+y,z)dy=\frac{\l z^{\pm1}}2\cF(z)\cR_{\pm}f+\frac{\l}2\cG\cR_{\pm}f
\end{multline}
and
\begin{multline}\lb{115}
 \int_{-\frac12}^{\frac12}F(\pm\frac12+y,z)f(\pm\frac12+y,z)dy=\l z^{\pm1}\int_{-\frac12}^{\frac12}F(2y,z)f(\pm\frac12+y,z)dy+\\
 \int_{-\frac12}^{\frac12}f(\pm\frac12+y,z)dy=\frac{\l z^{\pm1}}2\cF(z)\cR_{\pm}f+\int_{-\frac12}^{\frac12}f(\pm\frac12+y,z)dy,
\end{multline}
that after taking the sum of two terms with different signs $-$ and $+$ lead to
\[\lb{116}
 \cG(z)=\l\cF(z)\frac{z\cR_++z^{-1}\cR_-}2+\l\cG(z)\frac{\cR_++\cR_-}2,\ \ \cF(z)=\l\cF(z)\frac{z\cR_++z^{-1}\cR_-}2+\cJ.
\]
Formulas \er{116} allow us to express $\cF$ and $\cG$ explicitly
\[\lb{117}
 \cG(z)=\l\cF(z)\frac{z\cR_++z^{-1}\cR_-}2\lt(1-\l\frac{\cR_++\cR_-}2\rt)^{-1},\ \ \cF(z)=\cJ\lt(1-\l\frac{z\cR_++z^{-1}\cR_-}2\rt)^{-1}.
\]
Substituting the second formula of \er{117} into the first formula of \er{117} and remembering that $\cG$ is a free term in Taylor-Laurent $z$-series of $\cG(z)$, see \er{005} and similar arguments above \er{108}, we obtain finally \er{007}. Note that $\|z\cR_++z^{-1}\cR_-\|\le2$ in both spaces $C$ and $L^2$, when $|z|=1$. For the space of continuous functions it is obvious. For $L^2$, it follows from the definition \er{006} as
\begin{multline}\lb{118}
 \|zf(\frac{x+1}2)+z^{-1}f(\frac{x-1}2)\|_{L^2(-1,1)}^2\le\|zf(\frac{x+1}2)+
 z^{-1}f(\frac{x-1}2)\|_{L^2(-1,1)}^2+\\
 \|zf(\frac{x+1}2)-z^{-1}f(\frac{x-1}2)\|_{L^2(-1,1)}^2=2\|f(\frac{x+1}2)\|_{L^2(-1,1)}^2+2\|f(\frac{x-1}2)\|_{L^2(-1,1)}^2=\\
 4\|f(x)\|^2_{L^2(0,1)}+4\|f(x)\|^2_{L^2(-1,0)}=4\|f(x)\|^2_{L^2(-1,1)},
\end{multline}
where we use the fact that $|z|=1$.

Let us compute $\cG(z)e^{i\s x}$ for $\s\in\R$. We have
\begin{multline}\lb{119}
 \lt(1-\l\frac{z\cR_++z^{-1}\cR_-}2\rt)^{-1}e^{i\s x}=e^{i\s x}+\sum_{n=1}^{\iy}\frac{\l^n}{2^n}(z\cR_++z^{-1}\cR_-)^ne^{i\s x}=\\
 e^{i\s x}+
 \sum_{n=1}^{\iy}\frac{\l^n}{2^n}\prod_{j=1}^n(ze^{\frac{i\s}{2^j}}+z^{-1}e^{-\frac{i\s}{2^j}})\cdot e^{\frac{i\s x}{2^n}}.
\end{multline}
In particular,
\[\lb{119a}
 \lt(1-\l\frac{\cR_++\cR_-}2\rt)^{-1}e^{i\s x}=e^{i\s x}+\sum_{n=1}^{\iy}\l^n\prod_{j=1}^n\cos\frac{\s}{2^j}\cdot e^{\frac{i\s x}{2^n}}=\sin\s\sum_{n=0}^{\iy}\frac{\l^n}{2^n\sin\frac{\s}{2^n}}e^{\frac{i\s x}{2^n}}.
\]
Using \er{119} and \er{119a}, we obtain
\begin{multline}\lb{120}
 \lt(1-\l\frac{z\cR_++z^{-1}\cR_-}2\rt)^{-1}\l\frac{z\cR_++z^{-1}\cR_-}2\lt(1-\l\frac{\cR_++\cR_-}2\rt)^{-1}e^{i\s x}=\\
 \sin{\s}\sum_{m=1}^{\iy}\frac{\l^m}{2^m}\prod_{j=1}^m(e^{\frac{i\s}{2^{n+j}}}+e^{-\frac{i\s}{2^{n+j}}})\sum_{n=0}^{\iy}\frac{\l^n}{2^n\sin\frac{\s}{2^n}}
  e^{\frac{i\s x}{2^{n+m}}}=\\
  \sin{\s}\sum_{m=1}^{\iy}\sum_{n=0}^{\iy}\frac{\l^{m+n}\prod_{j=1}^m(e^{\frac{i\s}{2^{j+n}}}+e^{-\frac{i\s}{2^{j+n}}})e^{\frac{i\s x}{2^{m+n}}}}{2^{m+n}\sin\frac{\s}{2^n}}
.
\end{multline}
Let us define
\[\lb{121}
 C_m(\s):=\oint_{|z|=1}\prod_{j=1}^m(ze^{\frac{i\s}{2^j}}+z^{-1}e^{-\frac{i\s}{2^j}})\frac{dz}{2\pi i z}=2^m\int_{-\pi}^{\pi}\prod_{j=1}^m\cos(\vp+\frac{\s}{2^j})\frac{d\vp}{2\pi}.
\]
On the other hand, due to the first formula \er{121}, $C_m(\s)$ is a free term (without $z^j$ multipliers) in the expanded product $\prod_{j=1}^m(ze^{\frac{i\s}{2^j}}+z^{-1}e^{-\frac{i\s}{2^j}})$. Thus,
\begin{multline}\lb{122}
 C_m(\s)=\sum_{{\bf y}\in\{-1,1\}^m_0}e^{\sum_{j=1}^m\frac{i\s y_j}{2^j}}=\sum_{{\bf y}\in\{-1,1\}^m_0}\cos{\sum_{j=1}^m\frac{\s y_j}{2^j}},\ \ \ where\\ 
 \{-1,1\}^m_0=\{{\bf y}=(y_j)_{j=1}^m\in\{-1,1\}^m:\ \sum_{j=1}^my_j=0\}.
\end{multline}
Now, using already proved \er{007} along with \er{005}, \er{006}, and  \er{120}, \er{121}, we obtain
\[\lb{123}
 \int_{-1}^1U(x)e^{i\s x}dx=\frac{2\sin\s}{\s}\sum_{n=0}^{\iy}\frac{\l^n}{\sin\frac{\s}{2^n}}\sum_{m=1}^{\iy}\l^mC_m(\frac{\s}{2^n})\sin\frac{\s}{2^{n+m}}
\]
that gives \er{008}. Consider the special case $\s\in\pi\Z$, where \er{123} can be simplified by eleminating $\sin\s/\sin\frac{\s}{2^n}=\frac00$ singularities. Let $\s=2^pq\pi$ with non-negative integer $p$ and odd $q$. Consider the limit $\s\to2^pq\pi$. Eleminating $\frac{\sin\s}{\sin\frac{\s}{2^n}}$ in \er{123} and using \er{121}, we obtain
\begin{multline}\lb{124}
 \int_{-1}^1U(x)e^{i2^pq\pi x}dx=\frac{2}{\pi 2^pq}	\sum_{n=0}^{p}(-1)^{\d_{np}-\d_{p0}}(2\l)^n\sum_{m=1}^{\iy}\l^mC_m(\frac{\pi2^pq}{2^n})\sin\frac{\pi2^pq}{2^{n+m}}=\\
 \frac{2}{\pi 2^pq}	\sum_{n=0}^{p}(-1)^{\d_{np}-\d_{p0}}(2\l)^n\sum_{m=1+p-n}^{\iy}\l^mC_m(\frac{\pi2^pq}{2^n})\sin\frac{\pi2^pq}{2^{n+m}}=[n:=p-n]=\\
 \frac{2(2\l)^p}{\pi 2^pq}	\sum_{n=0}^{p}(-1)^{\d_{n0}-\d_{p0}}(2\l)^{-n}\sum_{m=1+n}^{\iy}\l^mC_m(\frac{\pi2^pq}{2^{p-n}})\sin\frac{\pi2^pq}{2^{p-n+m}}=\\
  \frac{2\l^p}{\pi q}	\sum_{n=0}^{p}(-1)^{\d_{n0}-\d_{p0}}(2\l)^{-n}\sum_{m=1+n}^{\iy}\l^mC_m(\pi2^nq)\sin\frac{\pi2^nq}{2^{m}}=[m:=m-n]=\\
  \frac{2\l^p}{\pi q}	\sum_{n=0}^{p}(-1)^{\d_{n0}-\d_{p0}}(2\l)^{-n}\sum_{m=1}^{\iy}\l^{m+n}C_{m+n}(\pi2^nq)\sin\frac{\pi q}{2^m}=\\
  \frac{2\l^p}{\pi q}	\sum_{n=0}^{p}(-1)^{\d_{n0}-\d_{p0}}2^{-n}\sum_{m=1}^{\iy}\l^{m}C_{m+n}(\pi2^nq)\sin\frac{\pi q}{2^m}=\\
  \frac{2\l^p}{\pi q}\int_{-\pi}^{\pi}	\sum_{n=0}^{p}(-1)^{\d_{n0}-\d_{p0}}\sum_{m=1}^{\iy}\l^m2^m\sin\frac{\pi q}{2^m}\prod_{j=1}^{n+m}\cos(\vp+\frac{\pi2^nq}{2^j})\frac{d\vp}{2\pi}=\\
  \frac{2\l^p}{\pi q}\int_{-\pi}^{\pi}	\sum_{n=0}^{p}(-1)^{1-\d_{p0}}\cos^n\vp\sum_{m=1}^{\iy}\l^m2^m\sin\frac{\pi q}{2^m}\prod_{j=1}^{m}\cos(\vp+\frac{\pi q}{2^j})\frac{d\vp}{2\pi}=\\
  \frac{(-1)^{1-\d_{p0}}2\l^p}{\pi q}\sum_{m=1}^{\iy}\l^m2^m\sin\frac{\pi q}{2^m}\int_{-\pi}^{\pi}\frac{(1-\cos^{p+1}\vp)\prod_{j=1}^{m}\cos(\vp+\frac{\pi q}{2^j})}{1-\cos\vp}\frac{d\vp}{2\pi}
\end{multline}
that give \er{009a}.
{\section{Examples}\lb{sec3}}
In this section we compute polynomials mentioned in Theorem \ref{T1} explicitly. 

\subsection{Integrals $\int_{-1}^{1}x^{A}U(x)dx$.} We use the notations \er{101}, \er{102}. Introduce
\[\lb{207}
 F_0(z)=\int_{-1}^1F(y,x)dx,\ \ \ G_0(z)=\int_{-1}^1G(x,z)dx.
\]
Integrating each of the equations in \er{105} and then taking their sum, we obtain
\[\lb{208}
 F_0(z)=2+\lt(\frac{\l z}2+\frac{\l}{2z}\rt)F_0(z)\ \Rightarrow\ F_0(z)=\frac{4z}{2z-\l z^2-\l}.
\]
Applying the same arguments as in \er{208} to \er{106}, we obtain also
\[\lb{209}
 G_0(z)=\frac{\l}{1-\l}\lt(\frac{z}2+\frac1{2z}\rt)F_0(z)=\frac{2\l(z^2+1)}{(1-\l)(2z-\l z^2-\l)}.
\] 
Let us denote
\[\lb{210}
 F_n(z)=\int_{-1}^{1}x^{n}F(x,z)dx,\ \ \ G_n(z)=\int_{-1}^{1}x^{n}G(x,z)dx,\ \ \ n\in\N.
\]
Again, using \er{105}, we get
\[\lb{211}
 \int_{0}^1F(x,z)x^ndx=\int_{-\frac12}^{\frac12}F(\frac12+y,z)(\frac12+y)^ndy=\frac1{n+1}+\frac{\l z}{2^{n+1}}\int_{-1}^1F(x,z)(1+x)^ndx,
\]
and
\[\lb{212}
 \int_{-1}^0F(x,z)x^ndx=\frac{(-1)^n}{n+1}+\frac{\l}{2^{n+1}z}\int_{-1}^1F(x,z)(-1+x)^ndx,
\]
which leads to 
\[\lb{213}
 F_n(z)=\frac{1+(-1)^n}{n+1}+\frac{\l}{2^{n+1}}\sum_{j=0}^{n}(z+(-1)^{n-j}z^{-1})\binom{n}{j}F_j(z)
\]
and
\[\lb{214}
 F_n(z)=\frac{\frac{1+(-1)^n}{n+1}+\l2^{-n-1}\sum_{j=0}^{n-1}(z+(-1)^{n-j}z^{-1})\binom{n}{j}F_j(z)}{1-\l2^{-n-1}(z+z^{-1})}
\]
or
\[\lb{215}
 F_n(z)=\frac1{n+1}\frac{(1+(-1)^n)2^{n+1}z+\l(n+1)\sum_{j=0}^{n-1}(z^2+(-1)^{n-j})\binom{n}{j}F_j(z)}{2^{n+1}z-\l z^2-\l}.
\]
Rewriting \er{106} (see \er{105}) in the form
\[\lb{216}
 G(\pm\frac12+y,z)=F(\pm\frac12+y,z)-1+\l G(2y,z),\ \ \ y\in[-\frac12,\frac12]
\]
and applying the same arguments as in \er{211}-\er{213}, we can write
\[\lb{217}
 G_n(z)=F_n(z)-\frac{1+(-1)^n}{n+1}+\frac{\l}{2^{n+1}}\sum_{j=0}^{n}(1+(-1)^{n-j})\binom{n}{j}G_j(z)
\]
which leads to
\[\lb{218}
 G_n(z)=\frac{F_n(z)}{1-\l 2^{-n}}-\frac{1+(-1)^n}{(1-\l2^{-n})(n+1)}+\frac1{2^{n+1}-2\l}\sum_{j=0}^{n-1}(1+(-1)^{n-j})\binom{n}{j}G_j(z).
\]
Thus, using \er{215} and \er{218}, we can obtain all $F_n$, $G_n$ by induction. It is seen that they are rational functions of $z$ with integer coefficients. Moreover, if we write $G_n$ as a ratio of two polynomials $G_n(z)=P_n(z)/Q_n(z)$ then we may see from \er{215} that $Q_n$ is a product of some number of square polynomials $2^{k+1}z-\l z^2-\l$ which has explicit roots expressed in terms of square roots of linear expressions depending on $\l$. Thus, the integral, see \er{112} and \er{210},
\[\lb{219a}
 \int_{-1}^{1}x^AU(x)dx=\frac1{2\pi i}\oint_{|z|=1}\frac{G_A(z)}{z}dz
\]
can be computed explicitly by the Cauchy residue theorem. The result of computations is a combination of rational functions of $\l$ and square roots of these rational functions. Note that by Lemma \ref{L1} the lower and upper limits $-1$ and $1$ can be replaced by dyadic rational numbers. In this case we also have a closed-form expression for the integral written in terms of rational functions of $\l$ and square roots of them. We continue the analysis of closed-form expressions in the next subsection.

\subsection{Integral $\int_{-1}^{1}U(x) R(x)dx$, where $R(x)$ is rational.} Such integrals can be computed with the help of the following function
\[\lb{H001}
 H(w)=\int_{-1}^1\frac{U(x)}{w-x}dx.
\]
So, using $H$, its derivatives and results of the first subsection, we can try to find some expressions for $\int_{-1}^{1}U(x) R(x)dx$ with rational $R$. In fact, $H$ is a modified Hilbert transform of $U$. It is analytic for $w\in\C\sm[-1,1]$. Unfortunately, it is not seen to me how $H$ can be expressed in terms of known functions. Nevertheless, we can derive some functional equations for the functions related to $H$. Define
\[\lb{H002}
 H_1(z,w)=\int_{-1}^{1}\frac{F(x,z)}{w-x}dx,\ \ \ H_2(z,w)=\int_{-1}^{1}\frac{G(x,z)}{w-x}dx.
\]
Then
\[\lb{HU}
 H(w)=\frac1{2\pi i}\oint_{|z|=1}\frac{H_2(z,w)}{z}dz.
\]
Using \er{105}, we obtain
\begin{multline}\lb{H003}
 \int_{0}^{1}\frac{F(x,z)}{w-x}dx=\int_{-\frac12}^{\frac12}\frac{F(\frac12+y,z)}{w-\frac12-y}dy=\int_{-\frac12}^{\frac12}\frac{1+\l zF(2y,z)}{w-\frac12-y}dy=\\
 \int_{-\frac12}^{\frac12}\frac{1+\l zF(2y,z)}{2w-1-2y}d(2y)=
 \ln(w)-\ln(w-1)+\l zH_1(z,2w-1)
\end{multline}
and
\begin{multline}\lb{H004}
	\int_{-1}^{0}\frac{F(x,z)}{w-x}dx=\int_{-\frac12}^{\frac12}\frac{F(-\frac12+y,z)}{w+\frac12-y}dy=\int_{-\frac12}^{\frac12}\frac{1+\l z^{-1}F(2y,z)}{w+\frac12-y}dy=\\
	\int_{-\frac12}^{\frac12}\frac{1+\l z^{-1}F(2y,z)}{2w+1-2y}d(2y)=
	\ln(w+1)-\ln(w)+\l z^{-1}H_1(z,2w+1).
\end{multline}
Taking the sum of \er{H003} and \er{H004} we get
\[\lb{H005}
 H_1(z,w)=\ln\frac{1+w^{-1}}{1-w^{-1}}+\l zH_1(z,2w-1)+\l z^{-1}H_1(z,2w+1).
\]
Similarly, using \er{106}, we get also
\[\lb{H006}
 H_2(z,w)=\l zH_1(z,2w-1)+\l z^{-1}H_1(z,2w+1)+\l H_2(z,2w-1)+\l H_2(z,2w+1).
\]
Both functions $H_1$ and $H_2$ are analytic in the neighborhood of $w=\iy$, where they can be expanded into the Loran series. Using definitions \er{210}, \er{H002}, and identities \er{H005}, \er{H006}, we deduce that
\[\lb{H007}
 \sum_{n=0}^{+\iy}\frac{F_n(z)}{w^{n+1}}=\sum_{n=0}^{+\iy}\frac{1+(-1)^{n}}{(n+1)w^{n+1}}+ \sum_{n=0}^{+\iy}\frac{\l zF_n(z)}{(2w-1)^{n+1}}+ \sum_{n=0}^{+\iy}\frac{\l z^{-1}F_n(z)}{(2w+1)^{n+1}}
\]
and
\[\lb{H008}
 \sum_{n=0}^{+\iy}\frac{G_n(z)}{w^{n+1}}= \sum_{n=0}^{+\iy}\frac{\l zF_n(z)+\l G_n(z)}{(2w-1)^{n+1}}+ \sum_{n=0}^{+\iy}\frac{\l z^{-1}F_n(z)+\l G_n(z)}{(2w+1)^{n+1}}
\] 
or
\[\lb{H009}
 \sum_{n=0}^{+\iy}\frac{G_n(z)}{w^{n+1}}= \sum_{n=0}^{+\iy}\frac{\l G_n(z)}{(2w-1)^{n+1}}+ \sum_{n=0}^{+\iy}\frac{\l G_n(z)}{(2w+1)^{n+1}}+\sum_{n=0}^{+\iy}\frac{F_n(z)-\frac{1+(-1)^n}{n+1}}{w^{n+1}}.
\]
Now, extracting terms corresponding to $w^{-n-1}$ for $n+1\in\N$, we obtain the linear systems that determine $F_n$ and $G_n$. In fact, \er{H005}, \er{H006} or \er{H007}-\er{H009} is a somewhat compact form of \er{213} and \er{217}. Indeed, substituting
\[\lb{H010}
 \frac1{(2w\pm1)^n}=\sum_{k=0}^{+\iy}\frac{(k+n)!(\mp1)^k}{(2w)^{k+n+1}k!n!}
\]
into \er{H007} we obtain
\begin{multline}\lb{H011}
	\sum_{n=0}^{+\iy}\frac{F_n(z)}{w^{n+1}}=\sum_{n=0}^{+\iy}\frac{1+(-1)^n}{(n+1)w^{n+1}}+\sum_{n=0}^{+\iy}\l F_n(z)\sum_{k=0}^{+\iy}\frac{(k+n)!(z+z^{-1}(-1)^k)}{(2w)^{k+n+1}k!n!}=[k+n=m]=\\
	\sum_{n=0}^{+\iy}\frac{1+(-1)^n}{(n+1)w^{n+1}}+\sum_{n=0}^{+\iy}\l F_n(z)\sum_{m=n}^{+\iy}\frac{(z+z^{-1}(-1)^{m-n})m!}{(2w)^{m+1}(m-n)!n!}=\\ \sum_{n=0}^{+\iy}\frac{1+(-1)^n}{(n+1)w^{n+1}}+
	\sum_{m=0}^{+\iy}\frac{\l}{(2w)^{m+1}} \sum_{n=0}^{m}F_n(z)(z+z^{-1}(-1)^{m-n})\binom{m}{n}=\\
	\sum_{n=0}^{+\iy}\frac{1+(-1)^n}{(n+1)w^{n+1}}+
	\sum_{n=0}^{+\iy}\frac{\l}{(2w)^{n+1}} \sum_{m=0}^{n}F_m(z)(z+z^{-1}(-1)^{n-m})\binom{n}{m}.
\end{multline}
Introducing the following infinite matrices and vector columns
\begin{multline}\lb{H012}
 {\bf J}=\diag\lt(1-\frac{\l(z+z^{-1})}{2^{n+1}}\rt)_{n\ge0},\ \ {\bf H}=\lt(\frac{\l(z+(-1)^{n-m}z^{-1})\binom{n}{m}\d_{n>m}}{2^{n+1}}\rt)_{n,m\ge0},\\ {\bf a}=\lt(\frac{1+(-1)^n}{n+1}\rt)_{n\ge0},\ \ {\bf f}=(F_n(z))_{n\ge0}
\end{multline}
we may write \er{H011} as
\[\lb{H013}
 ({\bf J}-{\bf H}){\bf f}={\bf a}\ \ or\ \ ({\bf I}-{\bf J}^{-1}{\bf H}){\bf f}={\bf J}^{-1}{\bf a}\ \ or\ \ ({\bf I}-{\bf K}){\bf f}={\bf b}\ \ with\ \ 
\]
\begin{multline}\lb{H014}
{\bf K}={\bf J}^{-1}{\bf H}=\lt(\frac{\l\binom{n}{m}(z^2+(-1)^{n-m})\d_{n>m}}{2^{n+1}z-\l z^2-\l}\rt)_{n,m\ge0},\\ {\bf b}={\bf J}^{-1}{\bf a}=\lt(\frac{2^{n+1}(1+(-1)^n)z}{(n+1)(2^{n+1}z-\l z^2-\l)}\rt)_{n\ge0},
\end{multline}
where $\d_{n>m}=1$ if $n>m$ and $\d_{n>m}=0$ otherwise, and ${\bf I}$ is the identity matrix. Since ${\bf K}$ is strongly lower triangular matrix, we may write the solution of \er{H013} in the form
\[\lb{H015}
 {\bf f}={\bf b}+{\bf K}{\bf b}+{\bf K}^2{\bf b}+{\bf K}^3{\bf b}+....
\]
Using \er{H012}, \er{H014} and \er{H015}, namely the fact that ${\bf K}$ is a lower triangular matrix with a zero diagonal, we may express the elements of ${\bf f}$ through the elements of ${\bf K}$ and ${\bf b}$
\[\lb{H016}
 F_{n_0}=b_{n_0}+\sum_{j=1}^{n_0}\sum_{n_0>n_1>...>n_j\ge0}b_{n_j}\prod_{i=1}^j K_{n_{i-1},n_i},\ \ \ n_0\ge0.
\]
Applying the same arguments as in \er{H007}, \er{H011} to \er{H009}, we obtain
\[\lb{H017}
  \sum_{n=0}^{+\iy}\frac{G_n(z)}{w^{n+1}}=\sum_{n=0}^{+\iy}\frac{F_n(z)-\frac{1+(-1)^n}{n+1}}{w^{n+1}}+\sum_{n=0}^{+\iy}\frac{\l}{(2w)^{n+1}} \sum_{m=0}^{n}G_m(z)(1+(-1)^{n-m})\binom{n}{m}.
\]
By analogy with \er{H012}-\er{H014}, introducing the infinite matrices and vector columns
\begin{multline}\lb{H018}
	\wt{\bf J}=\diag\lt(1-\frac{\l}{2^{n}}\rt)_{n\ge0},\ \ \wt{\bf H}=\lt(\frac{\l(1+(-1)^{n-m})\binom{n}{m}\d_{n>m}}{2^{n+1}}\rt)_{n,m\ge0},\\ \wt{\bf a}=\lt(F_n(z)-\frac{1+(-1)^n}{n+1}\rt)_{n\ge0},\ \ {\bf g}=(G_n(z))_{n\ge0}
\end{multline}
we may write \er{H017} as
\[\lb{H019}
(\wt{\bf J}-\wt{\bf H}){\bf g}=\wt{\bf a}\ \ or\ \ ({\bf I}-\wt{\bf J}^{-1}\wt{\bf H}){\bf g}=\wt{\bf J}^{-1}\wt{\bf a}\ \ or\ \ ({\bf I}-\wt{\bf K}){\bf g}=\wt{\bf b}\ \ with\ \ 
\]
\begin{multline}\lb{H020}
	\wt{\bf K}=\wt{\bf J}^{-1}\wt{\bf H}=\lt(\frac{\l\binom{n}{m}(1+(-1)^{n-m})\d_{n>m}}{2^{n+1}-2\l }\rt)_{n,m\ge0},\\ \wt{\bf b}=\wt{\bf J}^{-1}\wt{\bf a}=\lt(\frac{2^{n}((n+1)F_n(z)-1-(-1)^n)}{(n+1)(2^{n}-\l)}\rt)_{n\ge0}.
\end{multline}
Since $\wt{\bf K}$ is strongly lower triangular matrix, we may write the solution of \er{H019} in the form
\[\lb{H021}
{\bf g}=\wt{\bf b}+\wt{\bf K}\wt{\bf b}+\wt{\bf K}^2\wt{\bf b}+\wt{\bf K}^3\wt{\bf b}+....
\]
Using \er{H018}, \er{H020} and \er{H021}, namely the fact that $\wt{\bf K}$ is a lower triangular matrix with a zero diagonal, we may express the elements of ${\bf g}$ through the elements of $\wt{\bf K}$ and $\wt{\bf b}$
\[\lb{H022}
G_{n_0}=\wt b_{n_0}+\sum_{j=1}^{n_0}\sum_{n_0>n_1>...>n_j\ge0}\wt b_{n_j}\prod_{i=1}^j\wt K_{n_{i-1},n_i},\ \ \ n_0\ge0.
\]
Finally, we have
\[\lb{H023}
 \int_{-1}^1U(x)x^Adx=\oint_{|z|=1}G_{A}(z)\frac{dz}{2\pi i z}.
\]
Denoting $f_n(z):=z^{-1}F_n(z)$, we obtain announced formulas after Corollary \ref{C1}. Let us compute $\int_{-1}^1x^2U(x)dx$. Using \er{H016}, we obtain
\[\lb{H024}
 f_0(z)=\frac4{Q_0(z)},\ \ \ f_2(z)=\frac{16}{3Q_2(z)}+\frac{4\l(z^2+1)}{Q_0(z)Q_2(z)}+\\
 \frac{8\l^2(z^2-1)^2}{Q_0(z)Q_1(z)Q_2(z)},
\]
where quadratic polynomials
\[\lb{H025}
 Q_0(z)=2z-\l z^2-\l,\ \ \ Q_1(z)=4z-\l z^2-\l,\ \ \ Q_2(z)=8z-\l z^2-\l
\]
have roots
\[\lb{H026}
 z_0^{\pm}=\frac{1\pm\sqrt{1-\l^2}}{\l},\ \ \ z_1^{\pm}=\frac{2\pm\sqrt{4-\l^2}}{\l},\ \ \ 
 z_2^{\pm}=\frac{4\pm\sqrt{16-\l^2}}{\l}.
\]
Roots $z_j^{-}$ lie inside the unit ball, $z_j^{+}$ lie outside. Using Cauchy residue theorem, we compute
\begin{multline}\lb{H027}
\hat f_0:=\frac1{2\pi i}\int_{|z|=1}f_0(z)dz=\frac{4}{\l (z_0^+-z_0^-)}=\frac2{\sqrt{1-\l^2}},\ \ \ \hat f_2:=\frac1{2\pi i}\int_{|z|=1}f_2(z)dz=\\
\frac{16}{3\l(z_2^+-z_2^-)}+\frac{4\l(z_0^++z_0^-)}{\l(z_0^+-z_0^-)(8-\l(z_0^++z_0^-))}+\frac{4\l(z_2^++z_2^-)}{\l(z_2^+-z_2^-)(2-\l(z_2^++z_2^-))}+\\
\frac{8\l^2(z_0^+-z_0^-)^2}{\l(z_0^+-z_0^-)(4-\l(z_0^++z_0^-))(8-\l(z_0^++z_0^-))}+\frac{8\l^2(z_1^+-z_1^-)^2}{\l(z_1^+-z_1^-)(2-\l(z_1^++z_1^-))(8-\l(z_1^++z_1^-))}+\\
\frac{8\l^2(z_2^+-z_2^-)^2}{\l(z_2^+-z_2^-)(2-\l(z_2^++z_2^-))(4-\l(z_2^++z_2^-))}=\\
\frac{2}{3\sqrt{1-\l^2}}+\frac{4\sqrt{1-\l^2}}3-2\sqrt{4-\l^2}+\frac{2\sqrt{16-\l^2}}3,
\end{multline}
where we actively use the fact that $z_j^+z_j^-=1$. Now, using \er{H022}, \er{H023} and \er{H020}, we can compute
\begin{multline}\lb{H028}
	\int_{-1}^1x^2U(x)dx=\frac{4(3\hat f_2-2)}{3(4-\l)}+\frac{\hat f_0-2}{1-\l}\cdot\frac{2\l}{8-2\l}=\\
	\frac{2}{3(1-\l)}\lt(\frac1{\sqrt{1-\l^2}}-1\rt)+\frac{16\sqrt{1-\l^2}-24\sqrt{4-\l^2}+8\sqrt{16-\l^2}}{3(4-\l)}.
\end{multline}

\subsection{Integral $\int_{-1}^{1}U(x)^2dx$.} Let us define
\[\lb{219}
 F_1(z,w)=\int_{-1}^{1}F(x,z)F(x,w)dx,\ \ \ F_2(z,w)=\int_{-1}^{1}F(x,z)G(x,w)dx,
\]
\[\lb{220}
G_1(z,w)=\int_{-1}^{1}G(x,z)G(x,w)dx,\ \ \ G_2(z,w)=\int_{-1}^{1}G(x,z)F(x,w)dx.
\]
Using \er{105}, we obtain
\begin{multline}\lb{221}
 \int_{0}^{1}F(x,z)F(x,w)dx=\int_{-\frac12}^{\frac12}F(\frac12+y,z)F(\frac12+y,w)dy=\\
 1+\frac{\l z}{2}F_0(z)+\frac{\l w}{2}F_0(w)+\frac{\l^2zw}{2}F_1(z,w)
\end{multline}
and
\[\lb{222}
\int_{-1}^{0}F(x,z)F(x,w)dx=1+\frac{\l}{2z}F_0(z)+\frac{\l}{2w}F_0(w)+\frac{\l^2}{2zw}F_1(z,w),
\]
which leads to
\begin{multline}\lb{223}
 F_1(z,w)=\frac{2+\frac{\l(z+z^{-1})F_0(z)}2+\frac{\l(w+w^{-1})F_0(w)}2}{1-\frac{\l^2(zw+(zw)^{-1})}2}=
 \frac{4+\l(z+z^{-1})F_0(z)+\l(w+w^{-1})F_0(w)}{2-\l^2zw-\l^2(zw)^{-1}}=\\
 \frac{4zw+\l w(z^2+1)F_0(z)+\l z(w^2+1)F_0(w)}{2zw-(\l zw)^2-\l^2}.
\end{multline}
Similarly, using \er{105}, \er{106}, we obtain
\begin{multline}\lb{224}
\int_{0}^{1}F(x,z)G(x,w)dx=\int_{-\frac12}^{\frac12}F(\frac12+y,z)G(\frac12+y,w)dy=\frac{\l wF_0(w)}2+\frac{\l G_0(w)}2+\frac{\l^2zwF_1(z,w)}{2}+\\
\frac{\l^2zF_2(z,w)}{2},\ \ \int_{-1}^{0}F(x,z)G(x,w)dx=\frac{\l F_0(w)}{2w}+\frac{\l G_0(w)}2+\frac{\l^2F_1(z,w)}{2zw}+\frac{\l^2F_2(z,w)}{2z},
\end{multline}
which after taking the sum of two integrals, leads to
\begin{multline}\lb{225}
 F_2(z,w)=\frac{\frac{\l(w+w^{-1})F_0(w)}2+\l G_0(w)+\frac{\l^2(zw+(zw)^{-1})F_1(z,w)}2}{1-\frac{\l^2(z+z^{-1})}2}=\\
 \frac{\l(w+w^{-1})F_0(w)+2\l G_0(w)+\l^2(zw+(zw)^{-1})F_1(z,w)}{2-\l^2(z+z^{-1})}=\\
 \frac{\l z(w^2+1)F_0(w)+2\l zwG_0(w)+\l^2((zw)^2+1)F_1(z,w)}{(2z-(\l z)^2-\l^2)w}.
\end{multline}
Again, \er{105}, \er{106} allow us to obtain the similar formulas for $G_1(z,w)$. We skip the details that are the same as for \er{223} and \er{225}
\begin{multline}\lb{226}
 G_1(z,w)=\frac{\l^2}{2(1-\l^2)}((zw+(zw)^{-1})F_1(z,w)+(z+z^{-1})F_2(z,w)+(w+w^{-1})G_2(z,w))=\\
 =\frac{\l^2}{2zw(1-\l^2)}(((zw)^2+1)F_1(z,w)+w(z^2+1)F_2(z,w)+z(w^2+1)G_2(z,w)),
\end{multline}
where, see \er{219} and \er{220},
\[\lb{227}
 G_2(z,w)=F_2(w,z).
\]
Using \er{220}, \er{106} and \er{112}, we obtain finally
\[\lb{228}
 \int_{-1}^{1}U(x)^2dx=\frac1{(2\pi i)^2}\oint_{|w|=1}\oint_{|z|=1}\frac{G_1(z,w)}{zw}dzdw,
\]
where the integrand is a rational function of $z$ and $w$ with integer coefficients. Let us try to compute this integral by using the Cauchy residue theorem. We start from auxiliary expansions and integrals. We have
\begin{multline}\lb{229}
 2zw-(\l zw)^2-\l^2=-(\l z)^2\lt(w-\frac{c(\l^2)}{z}\rt)\lt(w-\frac{\wt c(\l^2)}{z}\rt)\ \ \ {\rm with}\\ c(\l):=\frac{1-\sqrt{1-\l^2}}{\l},\ \ \ \wt c(\l):=\frac{1+\sqrt{1-\l^2}}{\l}=\frac1{c(\l)},
\end{multline}
where, for $|\l|<1$ and $|z|=1$, the first root lies inside the unit ball $|w|<1$, the second one lies out the ball. Thus, we can compute the following integral explicitly
\begin{multline}\lb{230}
	I_1(\l):=\oint_{|z|=1}\frac{dz}{2\pi i}\oint_{|w|=1}\frac{(zw)^2+1}{zw}\cdot\frac{1}{2zw-(\l zw)^2-\l^2}\cdot\frac{dw}{2\pi i}=\\
	\oint_{|z|=1}\frac{-\l^{-2}z^{-1}dz}{2\pi i}+\oint_{|z|=1}\frac{2\l^{-2}}{-(\l z)^{2}(\wt c(\l^2)-c(\l^2))}\frac{dz}{2\pi i}=\frac1{\l^2\sqrt{1-\l^4}}-\frac1{\l^2},
\end{multline}
where we use the fact that $((zw)^2+1)/(zw)=2\l^{-2}$ when $2zw-(\l zw)^2-\l^2=0$. Next, note that
\[\lb{231}
 2z-\l z^2-\l=-\l(z-c(\l))(z-\wt c(\l)),
\]
where the first root lies inside the unit ball $|z|<1$ while the second one lies out, when $|\l|<1$. Thus,
\[\lb{232}
 I_2(\l):=\oint_{|z|=1}\frac{z^2+1}{z(2z-\l z^2-\l)}\cdot\frac{dz}{2\pi i}=-\l^{-1}+\frac{2\l^{-1}}{\l(\wt c(\l)-c(\l))}=\frac{1}{\l\sqrt{1-\l^2}}-\frac1{\l}.
\]
Using the same ideas as in \er{229}-\er{232}, we can compute 
\begin{multline}\lb{233}
I_3(\l):=\frac1{(2\pi i)^2}\oint_{|z|=1}\oint_{|w|=1}\frac{((zw)^2+1)(z^2+1)dwdz}{zw(2zw-(\l zw)^2-\l^2)(2z-\l z^2-\l)}=-\l^{-2}I_2(\l)+\\
\frac1{2\pi i}\oint_{|z|=1}\frac{2\l^{-2}(z^2+1)dz}{(\l z)^{2}z^{-1}(\wt c(\l^2)-c(\l^2))(2z-\l z^2-\l)}=\l^{-2}\lt(\frac1{\sqrt{1-\l^4}}-1\rt)I_2(\l).
\end{multline}
Changing the variables $z\leftrightarrow w$, we can compute also
\[\lb{234}
 I_4(\l):=\frac1{(2\pi i)^2}\oint_{|z|=1}\oint_{|w|=1}\frac{((zw)^2+1)(w^2+1)dwdz}{zw(2zw-(\l zw)^2-\l^2)(2w-\l w^2-\l)}=I_3(\l).
\]
Fubini's theorem along with \er{232} allows us to compute
\[\lb{235}
 I_5(\l):=\frac1{(2\pi i)^2}\oint_{|z|=1}\oint_{|w|=1}\frac{(w^2+1)(z^2+1)dwdz}{zw(2w-\l w^2-\l)(2z-(\l z)^2-\l^2)}=I_2(\l)I_2(\l^2).
\]
Using \er{230} and \er{232}, we can compute explicitly the next integral
\begin{multline}\lb{236}
 I_6(\l):=\frac1{(2\pi i)^2}\oint_{|z|=1}\oint_{|w|=1}\frac{((zw)^2+1)(z^2+1)dwdz}{zw(2zw-(\l zw)^2-\l^2)(2z-(\l z)^2-\l^2)}=\\ \frac1{2\pi i}\oint_{|z|=1}\frac{I_1(\l)(z^2+1)dz}{z(2z-(\l z)^2-\l^2)}=I_1(\l)I_2(\l^2).
\end{multline}
Using the same ideas as in \er{230}, we can compute
\begin{multline}\lb{237}
	I_7(\l):=\frac1{(2\pi i)^2}\oint_{|z|=1}\oint_{|w|=1}\frac{((zw)^2+1)(z^2+1)^2dwdz}{zw(2zw-(\l zw)^2-\l^2)(2z-(\l z)^2-\l^2)(2z-\l z^2-\l)}=\\ \frac1{2\pi i}\oint_{|z|=1}\frac{I_1(\l)(z^2+1)^2dz}{z(2z-(\l z)^2-\l^2)(2z-\l z^2-\l)}=\l^{-3}I_1(\l)+\\
	\frac{(2\l^{-2})^2I_1(\l)}{(2\l^2\l^{-2}\sqrt{1-\l^4})(2-2\l^{-1})}+\frac{(2\l^{-1})^2I_1(\l)}{(2-2\l)(2\l\l^{-1}\sqrt{1-\l^2})}=\\
	\l^{-3}\lt(1-\frac1{(1-\l)\sqrt{1-\l^4}}+\frac{\l}{(1-\l)\sqrt{1-\l^2}}\rt)I_1(\l).
\end{multline}
Let us compute another two integrals that are a little bit more complex
\begin{multline}\lb{238}
	I_8(\l,z):=\frac1{2\pi i}\oint_{|w|=1}\frac{((zw)^2+1)(w^2+1)dw}{w(2zw-(\l zw)^2-\l^2)(2w-\l w^2-\l)}=\l^{-3}+\\
	\frac{2\l^{-2}z(\l^{-4}z^{-2}(1-\sqrt{1-\l^4})^2+1)}{(\l z)^{2}\cdot2\l^{-2}z^{-1}\sqrt{1-\l^4}(2\l^{-2}z^{-1}(1-\sqrt{1-\l^4})-\l(\l^{-4}z^{-2}(1-\sqrt{1-\l^4})^2+1))}+\\
	\frac{(z^2\l^{-2}(1-\sqrt{1-\l^2})^2+1)2\l^{-1}}{(2z\l^{-1}(1-\sqrt{1-\l^2})-(\l z)^2\l^{-2}(1-\sqrt{1-\l^2})^2-\l^2)2\l\l^{-1}\sqrt{1-\l^2}}=\\
	\l^{-3}+\frac{z^2+c(\l^2)^2}{\l^3\sqrt{1-\l^4}(2\l^{-1}c(\l^2)z-c(\l^2)^2-z^2)}+\frac{z^2c(\l)^2+1}{\l\sqrt{1-\l^2}(2zc(\l)-(\l c(\l)z)^2-\l^2)}
\end{multline}
with $c$ and $\wt c$ defined in \er{229}.
The quadratic polynomials in \er{238} have the expansions
\begin{multline}\lb{240}
 2\l^{-1}c(\l^2)z-c(\l^2)^2-z^2=-(z-c(\l^2)c(\l))(z-c(\l^2)\wt c(\l)),\\
 2zc(\l)-(\l c(\l)z)^2-\l^2=-(\l c(\l))^2\lt(z-\frac{c(\l^2)}{c(\l)}\rt)\lt(z-\frac{\wt c(\l^2)}{c(\l)}\rt).
\end{multline}
Both roots of the first polynomial lie inside the ball $|z|<1$, the only first root of the second polynomial lies inside the ball $|z|<1$. Now, using \er{238}-\er{240}, we compute
\begin{multline}\lb{241}
	I_8(\l):=\frac1{(2\pi i)^2}\oint_{|z|=1}\oint_{|w|=1}\frac{((zw)^2+1)(z^2+1)(w^2+1)dwdz}{zw(2zw-(\l zw)^2-\l^2)(2w-\l w^2-\l)(2z-(\l z)^2-\l^2)}=\\
	\frac1{2\pi i}\oint_{|z|=1}\frac{(z^2+1)I_8(\l,z)dz}{z(2z-(\l z)^2-\l^2)}=\frac{I_2(\l^2)}{\l^3}+\frac1{\l^5\sqrt{1-\l^4}}+\\ 
	\frac{(\wt c(\l^2)^2+1)(\wt c(\l^2)^2+c(\l^2)^2)}{\l^3\sqrt{1-\l^4}\wt c(\l^2)\l^{2}(\wt c(\l^2)-c(\l^2))(2\l^{-1}c(\l^2)\wt c(\l^2)-c(\l^2)^2-\wt c(\l^2)^2)}+\frac1{\l^5\sqrt{1-\l^2}}+\\
	\frac{(c(\l^2)^2+1)(c(\l^2)^2c(\l)^2+1)}{\l\sqrt{1-\l^2}c(\l^2)\l^2(\wt c(\l^2)-c(\l^2))(2c(\l^2)c(\l)-(\l c(\l)c(\l^2))^2-\l^2)}+\\
	\frac{(\frac{c(\l^2)^2}{c(\l)^2}+1)(c(\l^2)^2+1)}{\l\sqrt{1-\l^2}\frac{c(\l^2)}{c(\l)}(2\frac{c(\l^2)}{c(\l)}-\frac{\l^2c(\l^2)^2}{c(\l)^2}-\l^2)\l c(\l)^2(\frac{\wt c(\l^2)}{c(\l)}-\frac{c(\l^2)}{c(\l)})}
	=\\
	\frac2{\l^5\sqrt{1-\l^4}}-\frac1{\l^5}+\frac{2-\l^4}{\l^5(1-\l^4)(\l^3-2+\l^4)}+\frac1{\l^5\sqrt{1-\l^2}}+\\
	\frac{1+\sqrt{1-\l^2}\sqrt{1-\l^4}}{\l^5\sqrt{1-\l^2}\sqrt{1-\l^4}(\l-1-\sqrt{1-\l^2}\sqrt{1-\l^4})}+\\
	\frac{1-\sqrt{1-\l^2}\sqrt{1-\l^4}}{\l^5\sqrt{1-\l^2}\sqrt{1-\l^4}(\l-1+\sqrt{1-\l^2}\sqrt{1-\l^4})}=\\
	\frac2{\l^5\sqrt{1-\l^4}}-\frac1{\l^5}+\frac{2-\l^4}{\l^5(1-\l^4)(\l^3-2+\l^4)}+\frac1{\l^5\sqrt{1-\l^2}}+\\ 
	\frac{2(1-\l^3-\l^4)}{\l^5\sqrt{1-\l^2}\sqrt{1-\l^4}(\l^3-2+\l^4)}.
\end{multline}
Now, using \er{223}, \er{230}, and \er{232}-\er{234}, we compute
\begin{multline}\lb{242}
 A_1(\l):=\frac1{(2\pi i)^2}\oint_{|z|=1}\oint_{|w|=1}\frac{((zw)^2+1)F_1(z,w)dwdz}{(zw)^2}=\\
 4I_1(\l)+4\l\l^{-2}I_3(\l)+4\l I_4(\l)=
 \frac4{\l^2}\lt(1-\frac1{\sqrt{1-\l^4}}\rt)\lt(1-\frac2{\sqrt{1-\l^2}}\rt).
\end{multline}
Using \er{225}, \er{223}, \er{235}-\er{237}, and \er{241}, we compute also
\begin{multline}\lb{243}
	A_2(\l):=\frac1{(2\pi i)^2}\oint_{|z|=1}\oint_{|w|=1}\frac{w(z^2+1)F_2(z,w)dwdz}{(zw)^2}=4\l I_5(\l)+\frac{4\l^2}{1-\l}I_5(\l)+\\
	4\l^2I_6(\l)+4\l^3I_7(\l)+4\l^3I_8(\l)=\frac4{\l^2(1-\l)}\lt(\frac1{\sqrt{1-\l^4}}-1\rt)\lt(\frac{1+\l}{\sqrt{1-\l^2}}-1-\frac{\l}{\sqrt{1-\l^4}}\rt)+\\
	\frac4{\l^2}\lt(	\frac2{\sqrt{1-\l^4}}-1+\frac{2-\l^4}{(1-\l^4)(\l^3-2+\l^4)}+\frac1{\sqrt{1-\l^2}}+
	\frac{2(1-\l^3-\l^4)}{\sqrt{1-\l^2}\sqrt{1-\l^4}(\l^3-2+\l^4)}\rt).
\end{multline}
Noting that
\[\lb{244}
 A_3(\l):=\frac1{(2\pi i)^2}\oint_{|z|=1}\oint_{|w|=1}\frac{z(w^2+1)G_2(w,z)dwdz}{(zw)^2}=A_2(\l),
\]
see \er{243} and \er{227}, we can compute \er{228}
\begin{multline}\lb{245}
 \int_{-1}^1 U(x)^2dx=\frac{\l^2}{2(1-\l^2)}(A_1(\l)+A_2(\l)+A_3(\l))=\frac{\l^2}{2(1-\l^2)}(A_1(\l)+2A_2(\l))=\\
 \frac2{(1-\l)(1-\l^2)}\lt(\frac1{\sqrt{1-\l^4}}-1\rt)\lt(-3+\l+\frac{4}{\sqrt{1-\l^2}}-\frac{2\l}{\sqrt{1-\l^4}}\rt)+\frac4{1-\l^2}\cdot\\
 \lt(	\frac2{\sqrt{1-\l^4}}-1+\frac{2-\l^4}{(1-\l^4)(\l^3-2+\l^4)}+\frac1{\sqrt{1-\l^2}}+
 \frac{2(1-\l^3-\l^4)}{\sqrt{1-\l^2}\sqrt{1-\l^4}(\l^3-2+\l^4)}\rt)=\\
 \frac2{1-\l^2}\lt(\frac1{\sqrt{1-\l^4}}+\frac{4(1-(1+\l)\sqrt{1+\l^2})}{(1-\l)(\l^3-2+\l^4)}+\frac{1+\l}{1-\l}-\frac{2(1+\l)}{(1-\l)\sqrt{1-\l^2}}\rt)=\\
 \frac2{1-\l^2}\lt(\frac1{\sqrt{1-\l^4}}+\frac{4\l}{(1-\l)^2(1+(1+\l)\sqrt{1+\l^2})}+\frac{1+\l}{1-\l}-\frac{2(1+\l)}{(1-\l)\sqrt{1-\l^2}}\rt)=\\
 \frac2{(1-\l^2)(1-\l)^2}\lt(\frac{(1-\l)^2}{\sqrt{1-\l^4}}+\frac{4\l}{1+(1+\l)\sqrt{1+\l^2}}+1-\l^2-2\sqrt{1-\l^2}\rt),
\end{multline}
where we use \er{226} and \er{242}-\er{244}.

{\section{Perspectives}\lb{sec4}}

Let us discuss other integrals that can be computed explicitly or can be reduced to integrals of some standard algebraic functions. For example, using integration by parts, we can compute integrals $\int x^A \wt U(x) dx$, where $\wt U(x)=\int_0^xU(y)dy$ is already continuous function. Another method is to use the dependence on parameter $\l$. We can take derivatives and integrals by $\l$ along with multiplication by $f(\l)$ to obtaint integrals $\int x^A V(x)dx$, where $V(x)$ is similar to $U(x)$ but with more complex loop's weights than $\l^n$, e.g. $R(n)\l^n$ with rational functions $R$.

The analysis done for $U$ can be directly generalized to $U_k$ defined by \er{003} and \er{002}, but with the condition $\sum_{j=n}^mx_j=k$. $k\in\Z$. This is possible because $G(x,z)=\sum_{k\in\Z}z^kU_k(x)$, see \er{102}. In particular, it is seen that for integrals of algebraic combinations of $U_k$, say $\int\sum_kU_k(x)^2dx$, the analysis is simpler than for single $\int U(x)^2dx=\int U_0(x)^2dx$, since, roughly speaking, we may use one $(2\pi i)^{-1}\int G_0(z)G_0(z^{-1})z^{-1}dz$ instead of two integrals in \er{228}.

Another interesting subject of study is a so-called multivariate loop counting functions
\[\lb{301}
 U(x_1,...,x_d)=\sum_{0\le n\le m<+\iy}\l^{m+1}L_{nm}(x_1)...L_{nm}(x_d).
\]
This subject of study is close to some open problem related to distribution of self-avoiding walks (SAW) in a multidimensional case. Our idea for a further research is to derive explicit formulas for integrals of polynomials of $U$ given in \er{301}. Then we will use this polynomials to approximate characteristic functions $\chi_{U<\ve}$ that, after integration, give us an approximation of some characteristics of distribution of SAW.

\begin{figure}
    \centering
    \begin{subfigure}[b]{0.99\textwidth}
        \includegraphics[width=\textwidth]{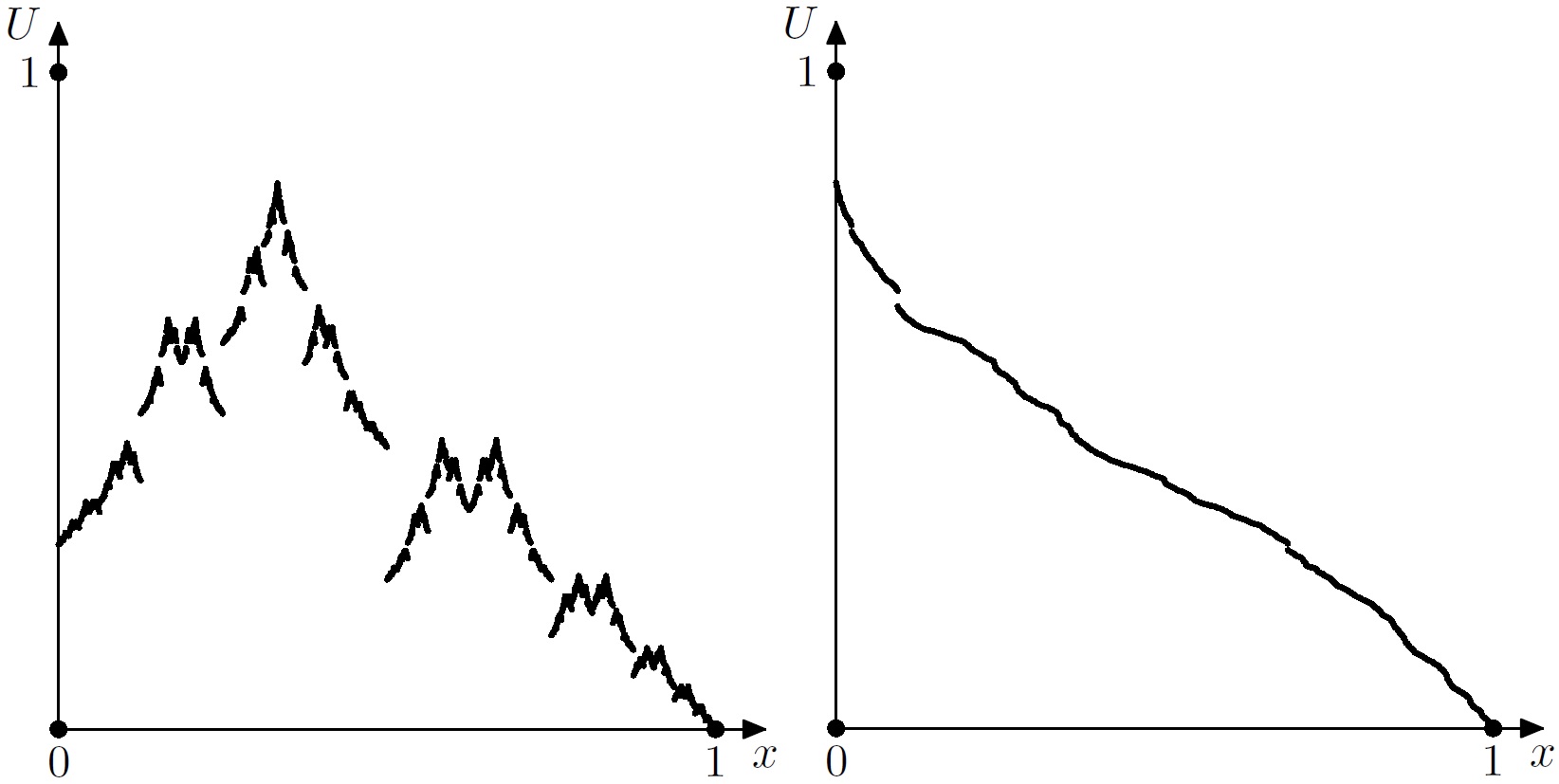}
        \caption{$\l=0.53$}
    \end{subfigure}  
    \begin{subfigure}[b]{0.99\textwidth}
        \includegraphics[width=\textwidth]{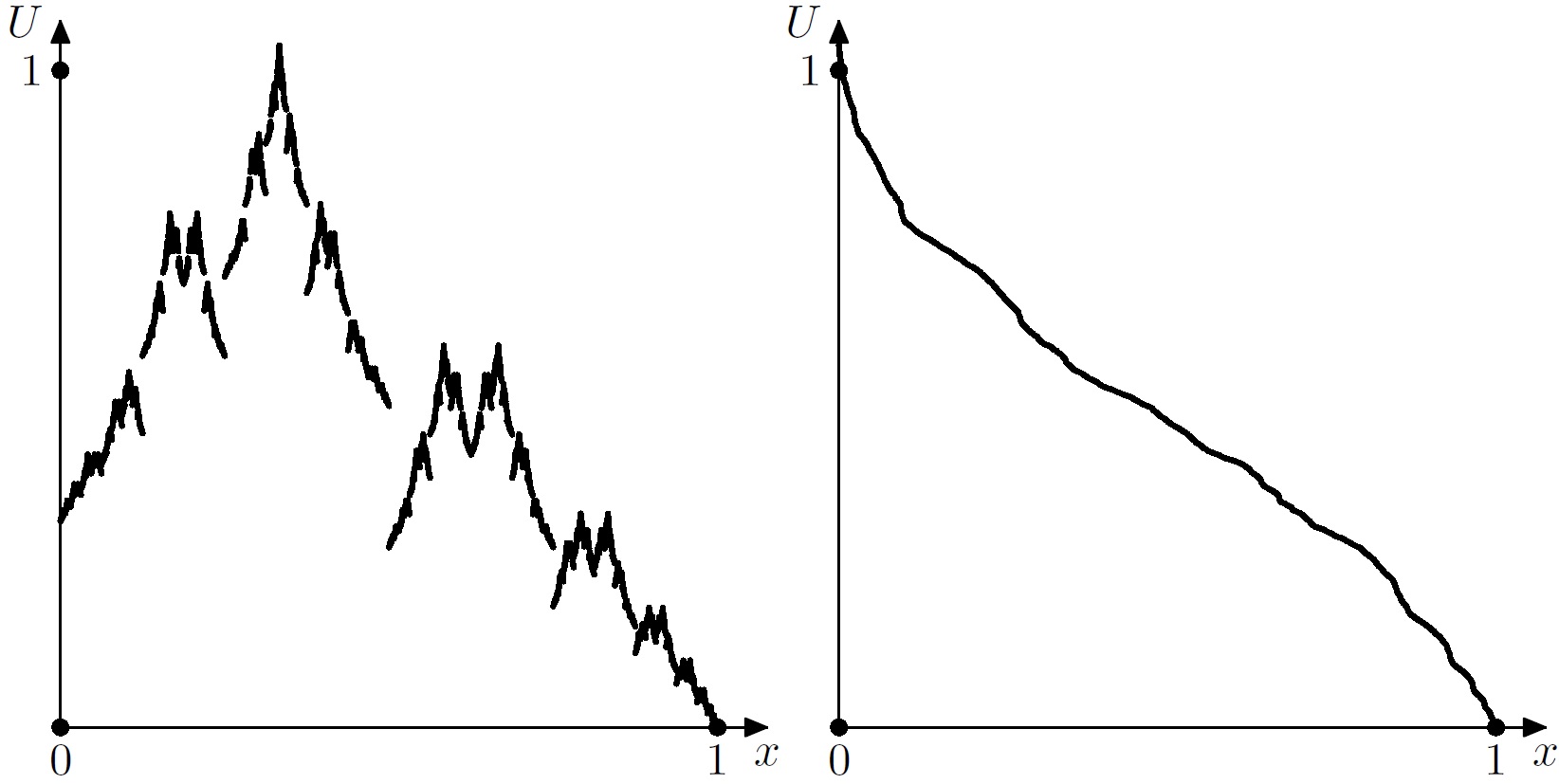}
        \caption{$\l=0.56$}
    \end{subfigure} 
    \caption{Plots of $U$ and its histogram at some specific values of $\l$. The histogram is a sorted array of values of $U(x_n)$ computed at the uniform net $x_n=n/N$, $1\le n\le N$ with $N=320001$.}\label{fig4}
\end{figure}

There are also certain expectations from the following topics. For polynomials $p$ we may define the linear operators
$$
 (\cA_{B}p)(\l)=\int_{-1}^1p(x)U(x)^Bdx,\ \ \ (\cB_{A}p)(\l)=\int_{-1}^1x^Ap(U(x))dx.
$$
The images of these operators are some functions analytic in $|\l|<1$. In particular, eigenvalue problems for $\cA_{B}$ and $\cB_{A}$ can be very helpful in understanding of their structure. Let us also briefly mention some problems related to the distribution of values of $U$. It is not difficult to see that for small real positive $\l\in(0,1)$ the image $U([0,1])$ is a sparse set similar to Cantor's sets. Computations show that there is $\l_0\approx0.6$ such that $U([0,1])$ is an interval $[U_{\rm min},U_{\rm max}]=[0,U(1/3)]=[0,\frac{\l^2}{(1-\l)(1-\l^2)}]$ without gaps. For the computation $U_{\rm max}$ we use the arguments from \cite{K3}: 
$$
	U_{\rm max}=U\lt(\frac13\rt)=\sum_{k=1}^{+\iy}\sum_{m,n\ge0,\ m-n=2k-1}\l^{m+1}=
	\sum_{k=1}^{+\iy}\sum_{m=2k-1}^{+\iy}\l^{m+1}=\sum_{k=1}^{+\iy}\frac{\l^{2k}}{1-\l}=\frac{\l^2}{(1-\l)(1-\l^2)}.
$$
Numerically estimating the points where the largest gap may appear, I firstly think that $\l_0=0.55496...$, which is a root of $x^3-2x^2-x+1$, but, perhaps, the situation is more complicated. The plots of $U$ and its histograms for $\l$ close to $\l_0$ are shown on Fig. \ref{fig4}.

\end{document}